\documentclass[twocolumn,showpacs,prb]{revtex4}
\usepackage{graphicx,dcolumn,bm,amsmath,amssymb,latexsym}

\begin{document}
  
\newcommand{\be}{\begin{equation}}
\newcommand{\ee}{\end{equation}}
\newcommand{\bea}{\begin{eqnarray}}
\newcommand{\eea}{\end{eqnarray}}
\newcommand{\br}{{\bf r}}

\title{Nonequilibrium mesoscopic conductance fluctuations}

\author{T.~Ludwig$^{1}$}
\author{Ya.~M.~Blanter$^{2}$}
\author{A.~D.~Mirlin$^{1,3,*}$}
\affiliation{
\mbox{$^{1}$ Institut f\"ur Nanotechnologie, Forschungszentrum
 Karlsruhe, 76021 Karlsruhe, Germany}\\
\mbox{$^{2}$ Kavli Institute of Nanoscience, Delft University of
  Technology, Lorentzweg 1, 2628 CJ Delft, The Netherlands}\\
\mbox{$^{3}$ Institut f\"ur Theorie der Kondensierten Materie,
 Universit\"at Karlsruhe, 76128 Karlsruhe, Germany}
}

\date{\today}

\begin{abstract}
We investigate the amplitude of mesoscopic fluctuations of the differential
conductance of a metallic wire at arbitrary bias voltage $V$. 
For non-interacting electrons, the variance $\langle \delta g^2\rangle$
increases with $V$. The asymptotic large-$V$ behavior is 
\mbox{$\langle \delta g^2\rangle \sim V/V_c$} (where \mbox{$eV_c=D/L^2$}
is the Thouless energy), in agreement with the earlier prediction by
Larkin and Khmelnitskii.
We find, however, that this asymptotics has a very small numerical
prefactor and sets in at very large $V/V_c$ only, which strongly
complicates its experimental observation.
This high-voltage behavior is preceded by a crossover regime,
\mbox{$V/V_c\lesssim 30\,$}, where the conductance variance increases by
a factor $\sim 3$ as compared to its value in the regime of universal
conductance fluctuations (i.e., at \mbox{$V\to 0$}).
We further analyze the effect of dephasing due to the
electron-electron scattering on $\langle \delta g^2\rangle$ at high
voltages. With the Coulomb interaction taken into account, the
amplitude of conductance fluctuations becomes a nonmonotonic
function of $V$. Specifically, $\langle \delta g^2\rangle$ drops as
$1/V$ for voltages \mbox{$V\gg gV_c\,$}, where $g$ is the
dimensionless conductance. In this regime, the conductance
fluctuations are dominated by quantum-coherent regions of the wire
adjacent to the reservoirs.

\end{abstract}


\maketitle

\section{Introduction}
\label{intro}

Universal conductance fluctuations
\cite{altshuler85,lee85,Lee_Stone_Fukuyama} (UCF) represent one of the
most famous manifestations of quantum-coherent electron transport in
mesoscopic conductors. Due to the quantum coherence of the electron
motion over the entire sample, the variance $\langle\delta g^2\rangle$
of the dimensionless (measured in units of $e^2/h$) conductance $g$ is
of order unity, with no dependence on the system size. The numerical
coefficient depends on the spatial dimensionality and on global
symmetries of the Hamiltonian. In particular, for a
quasi-one-dimensional system (a wire) and for preserved time-reversal
and spin-rotation symmetries one has
\mbox{$\langle\delta g^2\rangle=8/15\,$}. 

In a pioneering paper\cite{Larkin_Khmelnitskii} Larkin and
Khmelnitskii (LK) predicted an enhancement of mesoscopic
fluctuations of the differential conductance $g=dI/dV$ at high bias
voltages $V$. Specifically, they found that for \mbox{$V\gg V_c\,$},
where $eV_c=D/L^2$ is the Thouless energy, $D$ the diffusion constant,
and the $L$ the system size, the variance of the conductance increases
linearly with $V$, \mbox{$\langle\delta g^2\rangle\sim V/V_c\,$}.
This result implies, in particular, that regions of negative 
differential resistance should appear in the $IV$-characteristics
of a mesoscopic sample at sufficiently high voltages,
\mbox{$V/V_c\gtrsim g^2$}. 

While early measurements\cite{webb88,kaplan88,ralph93}
did not confirm the predicted enhancement of the conductance
fluctuations by the bias voltage, more recent experiments
\cite{schaefer96,weber00,Strunk,Terrier_PhD}
did observe an increase of  $\langle\delta g^2\rangle$
with $V$. This increase was interpreted as the linear behavior 
\mbox{$\langle\delta g^2\rangle\sim V/V_c$} predicted by LK. 

Inelastic scattering processes lead to dephasing, thus suppressing
interference phenomena like mesoscopic conductance fluctuations.
Indeed, it was found in all experiments that at sufficiently high
voltages the amplitude of conductance fluctuations starts to decrease
with $V$. However, no quantitative theory of the effect of Coulomb
interaction on the conductance fluctuations has been developed. LK 
\cite{Larkin_Khmelnitskii} characterized the effect of inelastic
scattering by a phenomenological inelastic out-scattering time
$\tau_{\rm in}$ and the corresponding length
$L_{\rm in}=\sqrt{D\tau_{\rm in}}\,$. They obtained a suppression
factor of $\langle\delta g^2\rangle$ of the form
\mbox{$\sim (L_{\rm in}/L)^7$} for the case
\mbox{$L_{\rm in}\ll L\,$}.
No microscopic theory for $L_{\rm in}$
(which would produce, in particular, its explicit dependence on $V$)
was presented in Ref.~\onlinecite{Larkin_Khmelnitskii}. The authors of
Refs.~\onlinecite{Strunk} and \onlinecite{Terrier_PhD} identified the $V$-dependent
dephasing length $L_\phi(V)$ as a relevant length scale induced by the
inelastic scattering. They further argued that for the case of Coulomb
interaction $L_\phi(V)$ can be obtained from the known
result\cite{Altshuler_Aronov,AAK} for $L_\phi(T)$ by a replacement 
\mbox{$T\to  V$}.  
They estimated the suppression factor of $\langle\delta g^2\rangle$ by
the inelastic processes as $(L_\phi(V)/L)^4$, at variance with LK. 
It was found experimentally in Ref.~\onlinecite{Strunk} that
$\langle\delta g^2\rangle$ decreases as $V^{-\gamma}$ at large $V$,
with \mbox{$\gamma = 1.28 \pm 0.12\,$}. The authors argued that the
electron-electron scattering is not sufficient to convert the LK
enhancement mechanism into such a fall-off, and ascribed this behavior
to the electron-phonon scattering.

The purpose of this paper is to develop a systematic theory of the
voltage dependence of conductance fluctuations of mesoscopic wires. 
In Sec.~\ref{non-int}
we calculate explicitly the dependence of  $\langle\delta g^2\rangle$
on $V/V_c$ for not too high voltages \mbox{$V\ll gV_c\,$}, when (as we
will see in Sec.~\ref{int}) the electron-electron scattering processes
do not play any role. While we confirm the LK asymptotic result
$\langle \delta g^2\rangle \sim V/V_c$ at \mbox{$V\gg V_c\,$}, we find
that the corresponding contribution has a very small numerical
prefactor. As a result, the LK asymptotic behavior sets in at very
large $V/V_c$ only and can hardly be detected experimentally.
This high voltage behavior is preceded by a crossover regime,
\mbox{$V/V_c\lesssim 30$}, where the conductance variance increases by
a factor $\sim 3$ as compared to its value in the regime of universal
conductance fluctuations (i.e., at \mbox{$V\to 0$}).
In contrast to the weak increase of the conductance fluctuations in
the LK regime, this strong enhancement of $\langle\delta g^2\rangle$
in the crossover regime can easily be verified experimentally.
In particular, we believe that it is this behavior that has been
observed in the most recent and detailed experimental
study.\cite{Strunk,Terrier_PhD}

In Sec.~\ref{finite-T} we consider the effect of finite
temperature, $T\gtrsim eV_c\,$. 
We show that the temperature does not affect the conductance
fluctuations at high voltages $eV\gg T$ (up to a constant offset) but
strongly suppresses them at lower voltages $eV\ll T\,$.

In Sec.~\ref{int} we analyze the effect of dephasing due to the
electron-electron interaction on $\langle \delta g^2\rangle$ at high
voltages. We show that the Coulomb interaction makes the amplitude of
conductance fluctuations a nonmonotonic function of $V$.
Specifically, we find that $\langle\delta g^2\rangle$ drops as $1/V$
for voltages \mbox{$V\gg gV_c\,$}, where $g$ is the dimensionless
conductance.
Our results demonstrate that in the non-equilibrium situation the
sample cannot be characterized by a single dephasing length, since the
latter is position-dependent and strongly increases near the contacts
with the reservoirs. Furthermore, we show that the conductance
fluctuations are dominated by quantum-coherent segments of the wire
adjacent to the reservoirs.
According to our results, the maximal rms conductance fluctuations
$\sqrt{\langle\delta g^2\rangle}$ are of the order of
$\sqrt{\smash[b]{g}}$ and occur at voltages of the order of $gV_c\,$.
Thus, the regions of negative differential resistance predicted by LK
in the high-voltage regime do not appear.

Sec.~\ref{compare-exp} is devoted to a comparison with experimental
data of Refs.~\onlinecite{Strunk} and \onlinecite{Terrier_PhD}.
Our results are summarized in Sec.~\ref{summary}. Technical details of
the calculations are presented in several Appendixes. 

\section{Non-interacting electrons}
\label{non-int}

We consider a quasi-one-dimensional metallic sample (a wire) of length
$L$ attached to two reservoirs with a voltage difference $V$.  
In order to calculate the conductance variance in the strongly
non-equilibrium regime, i.e.~at a large bias voltage $V$, 
we employ the Keldysh technique (see Ref.~\onlinecite{Keldysh}, for a
review see Ref.~\onlinecite{Rammer_Smith}).
In the Keldysh formalism the ensemble averaged current can be
represented with the help of the ensemble averaged diagonal Keldysh
function \mbox{$\langle G^K_\epsilon(x)\rangle\equiv\langle
G_\epsilon^K(x,x)\rangle=2\pi i\nu[2n_\epsilon(x)-1]\,$}, 
where $n_\epsilon$ is the electron distribution function,
\bea
\langle I_x\rangle&=&\hspace*{3mm}\parbox{4.2cm}{
\includegraphics[width=1.0\linewidth]{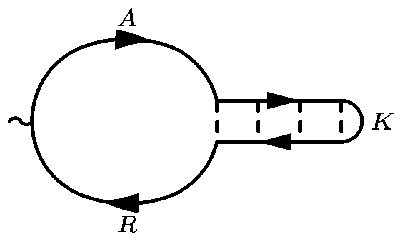}
}\nonumber\\
& &\nonumber\\
&=&\frac{e}{4\pi i}\,D\,\partial_x\int\!d\epsilon\:\big\langle G^K_\epsilon(x)\big\rangle\:\:.\label{e1}
\eea
Here $D$ is the diffusion constant, $\nu$ is the density of states,
$x$ is the coordinate along the wire,
$R$, $A$, and $K$ denote the retarded, advanced, and Keldysh Green's
functions respectively, and the wiggly line denotes the external
current vertex, with which an operator $e\hat{v}_x$ is associated.
The derivation of Eq.~(\ref{e1}) is outlined in Appendix~\ref{app-current}.
The diagrams for current fluctuations can be obtained by connecting
two current diagrams by impurity ladders in all possible ways.
We get the following six diagrams:
\bigskip

\includegraphics[width=0.95\linewidth]{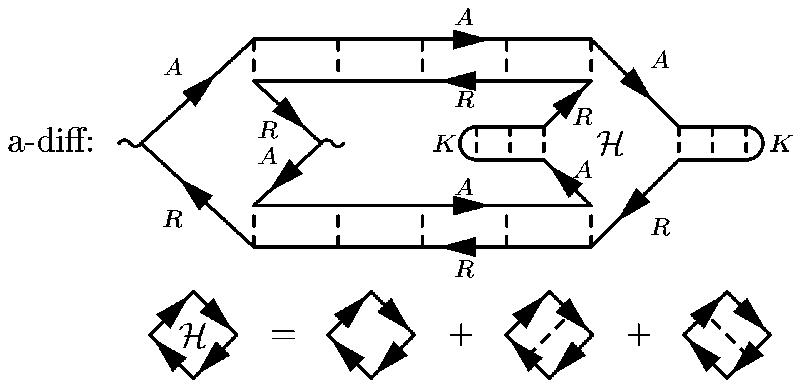}
\smallskip

\includegraphics[width=0.95\linewidth]{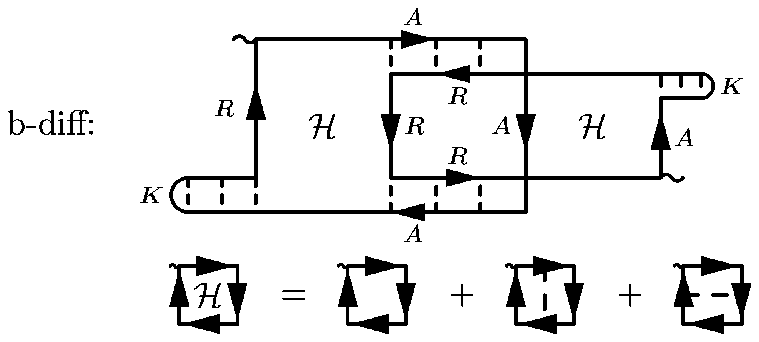}
\smallskip

\includegraphics[width=0.95\linewidth]{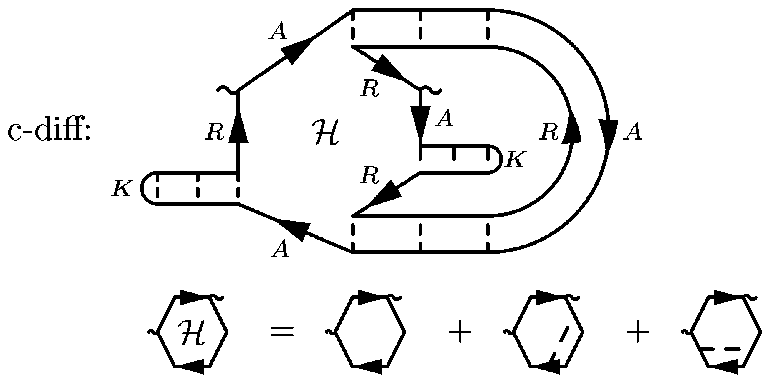}
\smallskip

\includegraphics[width=0.95\linewidth]{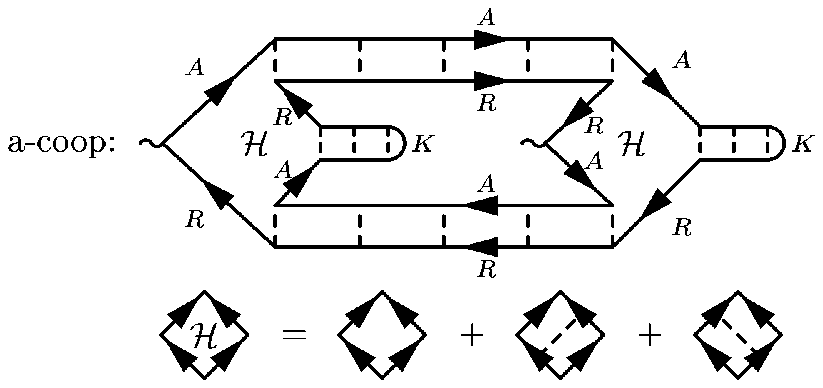}
\smallskip

\includegraphics[width=0.95\linewidth]{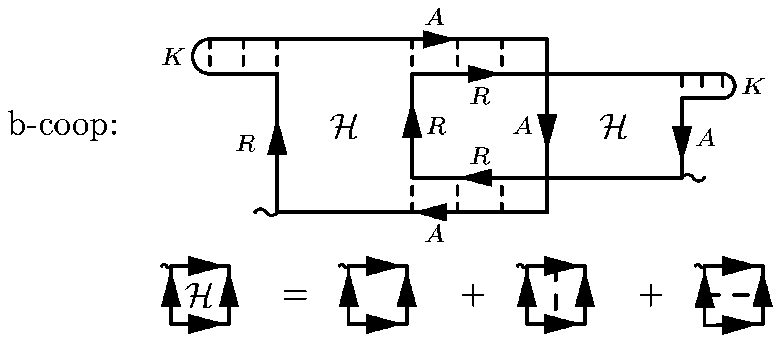}
\smallskip

\includegraphics[width=0.95\linewidth]{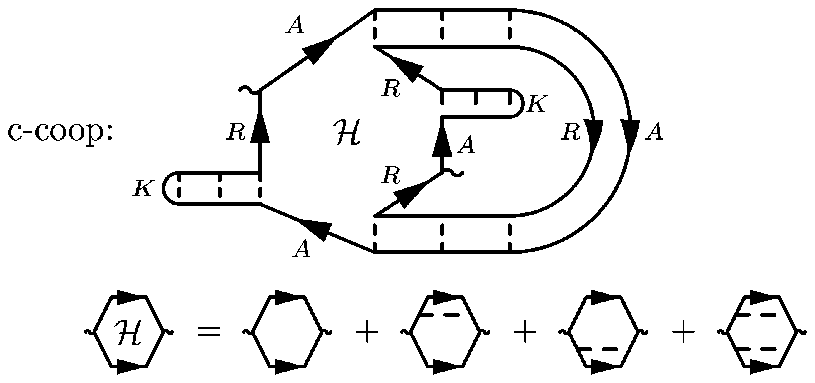}
\medskip

As usual in the impurity diagram technique, the diagrams consist of
electronic vertices (Hikami boxes) connected by diffusons and
Cooperons.
The Hikami boxes take into account the possibility of inserting
additional impurity lines connecting electronic Green functions of the
same type (retarded with retarded, or advanced with advanced) without
crossings. For the hexagonal vertices of the diagrams c-diff and
c-coop, only the insertions contributing to the leading (zeroth) order
in the external momenta are shown.
Evaluating all the diagrams (the technical details of the
calculation are presented in  Appendix~\ref{app-diagrams}), we get the
following result for the correlation function of mesoscopic
fluctuations of the current,\cite{lk-note}
\bea
\lefteqn{\big\langle\delta I(V_1)\,\delta I(V_2)\big\rangle =}\nonumber\\
&
&\!\!\!\!\!\!-\left(\frac{1}{2\pi\nu}\right)^2
\!\!\!\int\!\!d\epsilon_1d\epsilon_2\!\int\!\!\frac{dx_1dx_2}{L^2}
\frac{\partial}{\partial x_1}
\langle G^K_{\epsilon_1}(x_1)\rangle\frac{\partial}{\partial
  x_2}\langle G^K_{\epsilon_2}(x_2)\rangle\nonumber\\ 
&
&\qquad\times\left[\,2\left|
\Pi_{\epsilon_1-\epsilon_2}(x_1,x_2)\right|^2+{\rm 
    Re}\,\Pi^2_{\epsilon_1-\epsilon_2}(x_1,x_2)\,\right]. 
\label{II}
\eea
Here $\Pi$ is a rescaled diffusion propagator satisfying
the equation
\be
\left\{\partial_x^2 + \frac{i\omega}{D} + 
\frac{ie}{D}\Big[\phi_1(x)-\phi_2(x)\Big]\right\} 
\Pi_\omega(x,x^\prime) = -\delta(x-x^\prime)
\ee
with the boundary conditions that $\Pi=0$ if any of the coordinates
$x,x'$ is either $0$ or $L$,
and $\phi_{1,2}=(x/L)V_{1,2}$ are the potentials
corresponding to the bias voltages $V_{1,2}\,$.
Equation (\ref{II}) is written for spinless electrons and unbroken
time-reversal symmetry.
Alternatively, it is valid for spinful electrons with a strong
spin-orbit interaction, when only the spin-singlet channel
contributes. If the spin rotation symmetry is preserved,
Eq.~(\ref{II}) should be multiplied by a factor 4.
If time-reversal symmetry is broken, Eq.~(\ref{II}) should be
multiplied by an extra factor of $1/2\,$.
It is worth emphasizing  that, when the complete set of diagrams is
taken into account, they combine in such a way that only the spatial
derivative of the Keldysh function enters the result, Eq.~(\ref{II}).
This is because only the electrons in the energy window $V$ contribute
to the transport. This property, which does not hold for individual
diagrams but only for the sums of the diffuson or Cooperon diagrams,
can serve as a useful check of the calculation.

We assume first that the temperature of reservoirs is sufficiently low,
$T\ll eV_c$ (or, equivalently, the thermal diffusion length
$L_T=(D/T)^{1/2}$ is much larger than the system size), so that we can
set $T=0\,$. In other words, we will assume that at low voltage the
system is in the regime of universal conductance fluctuations.
The effect of a finite temperature, $T\gtrsim eV_c\,$, will be
analyzed in Sec.~\ref{finite-T}.

When a quasi-1d wire of length $L$ is connected to two perfect
reservoirs with different Fermi energies, the electron distribution
function in the wire will not be a Fermi function, but have a
double-step shape.
In the Keldysh formalism, this follows immediately from the kinetic
equation for the average Keldysh function,
\be
\label{kineq}
D\nabla^2\big\langle G^K_\epsilon(x)\big\rangle = 0
\ee
with the boundary conditions 
\be
\label{boundary}
\big\langle G^K_\epsilon(x)\big\rangle = -2\pi i\nu \times
\left\{
\begin{array}{cc}
1-2f(\epsilon)\:, &    x=0\\
1-2f(\epsilon-eV)\:, &\: x=L\:,
\end{array}
\right.
\ee
where $f$ is the Fermi function. 
This equation implies that the Keldysh
function is a linear function of the coordinate,
\be
\label{double-step}
\big\langle G^K_\epsilon(x)\big\rangle = -2\pi i\nu
\Big\{1-2f(\epsilon)+2{\textstyle\frac{x}{L}}\big[f(\epsilon)-
f(\epsilon-eV)\big]\Big\}\:.
\ee
This yields (including a factor $4$ to account for the spin)
\be
\big\langle\delta I(V_1)\,\delta I(V_2)\big\rangle =
16\,V_c^2\!\!\int\limits_0^{V_1/V_c}\!\!dz_1 
\!\!\int\limits_0^{V_2/V_c}\!\!dz_2\:\:\Xi_{z_1-z_2}\:,
\label{II_simplified}
\ee
where
\bea
\label{Xi_z}
\Xi_z&=&\int\limits_0^1 dy_1dy_2\,\left[2\left|\Pi_z(y_1,y_2)\right|^2
  +{\rm Re}\,\Pi^2_z(y_1,y_2)\right]\nonumber\\
&\equiv&{\rm Tr}\left[2\left|\Pi_z\right|^2+{\rm Re}\,\Pi^2_z\right]
\eea
and
\be
\big(\partial_y^2 + iz + i\alpha
  y\big)\,\Pi_z(y,y^\prime) = -\delta(y-y^\prime)\:.
\ee
Here we have introduced dimensionless variables \mbox{$y=x/L\,$},
\mbox{$z=\omega/eV_c\,$}, and \mbox{$\alpha=(V_1-V_2)/V_c\,$}.

To get the fluctuations of the differential conductance, we
differentiate Eq.~(\ref{II_simplified}) with respect 
to both voltages and set \mbox{$V_1=V_2\equiv V$}.
The result has the form (we set $T=0$ as discussed above)
\be
\left\langle\delta g\,\delta g\right\rangle=\left\langle\delta
  g\,\delta g\right\rangle_0+\left\langle\delta g\,\delta
  g\right\rangle_1+\left\langle\delta g\,\delta g\right\rangle_2\:, 
\ee
where
\bea
\big\langle\delta g(V)\,\delta
g(V)\big\rangle_0&=&16\:\Xi_0\big|_{\alpha=0}={8\over 15}\:,
\label{Xi_0} \\
\big\langle\delta g(V)\,\delta
  g(V)\big\rangle_1&=&32\!\!\!\int\limits_0^{V/V_c}\!\!\!dz\,
\frac{\partial}{\partial\alpha}\Xi_{z-\frac{V}{V_c}}\Big|_{\alpha=0}\,,
\label{Xi_1}\\  
\big\langle\delta g(V)\,\delta g(V)\big\rangle_2 &=&
\!-16\!\!\!\int\limits_0^{V/V_c}\!\!\!\!dz_1dz_2\frac{\partial^2}
{\partial\alpha^2}\Xi_{z_1-z_2}\Big|_{\alpha=0}.
\label{Xi_2}
\eea

Let us now discuss the physical meaning of the three contributions
(\ref{Xi_0}), (\ref{Xi_1}), and (\ref{Xi_2}) to the variance of the
differential conductance. Consider first the low-voltage (linear)
regime \mbox{$V \ll V_c\,$}. In this case, the only effect of a
voltage increment $dV$  is to change the chemical potential in the
reservoirs, and thus to make the states with energies in the window
$dV$ available for electron transmission. The corresponding 
conductance fluctuations are given by the term (\ref{Xi_0}),
reproducing the well-known UCF result,\cite{Lee_Stone_Fukuyama}
$\langle\delta g^2\rangle = 8/15\,$.
The two other terms vanish at low voltages and thus describe truly
non-equilibrium effects. Indeed, out of equilibrium the effect of the
applied voltage is not only to change the chemical potential in the
reservoirs, but also to alter the distribution of electric field in
the sample. The shift $dV$ of the voltage induces the variation
$d\phi(x)=(x/L)dV$ of the electrostatic potential, and this variation
affects all the electrons in the energy window $V$ contributing to the
current. This is the origin of the third term, Eq.~(\ref{Xi_2}). The
second one, Eq.~(\ref{Xi_1}), is the cross-term due to correlations
between the above two random contributions to $dI/dV$. 

The following comment is in order here. When considering the
conductance in terms of the scattering-theory formalism, one often
replaces the problem by that of non-interacting electrons with
different chemical potentials (or, equivalently, different
concentrations) in the reservoirs. While the contribution
$\left\langle\delta g\,\delta g\right\rangle_0$
is correctly reproduced in this way, the other two terms would be
lost. What is not taken into account by such a non-interacting
picture is that, for a system size much larger than the screening
length, the main effect of changing the \mbox{(electro-)}chemical
potential is not in altering the concentration but rather in changing
the electrostatic potential in the reservoirs and thus modifying the
potential profile in the wire. 

To evaluate the derivatives in Eqs.~(\ref{Xi_1}) and (\ref{Xi_2}),
we expand $\Pi$ up to second order in $\alpha$,
\be
\label{Pi_z}
\Pi_z = \Pi_z^{(0)} +i\alpha\Pi_z^{(0)} y\Pi_z^{(0)} -\alpha^2
\Pi_z^{(0)}y\Pi_z^{(0)}y\Pi_z^{(0)} + {\cal O}(\alpha^3)\:,
\ee
where \mbox{$\Pi_z^{(0)}\equiv\left(-\partial_y^2-{\rm
      i}z\right)^{-1}$}, and use the representation diagonalizing
$\Pi_z^{(0)}$, 
\be
\left(-\partial_y^2-{\rm
    i}z\right)=\sum\limits_{n>0}\lambda_n\,|n\rangle\langle n|\:,
\ee
where \mbox{$|n\rangle=\sqrt{2}\,{\rm sin}(n\pi y)\,$}
and \mbox{$\:\lambda_n=(n\pi)^2-iz\,$}. The matrix elements of $y$ in
this representation are given by
\bea
\langle n|y|m\rangle &=&
\frac{4\left[\left(-1\right)^{m+n}-1\right]m\,n}{\left(m-n\right)^2\,\left(m+n\right)^2\,\pi^2}\:\:,
\qquad n\neq m\nonumber\\
\langle n|y|n\rangle &=& \frac{1}{2}\:\:.
\eea

We first find the asymptotic behavior of $\langle\delta g^2\rangle$ in
the limit $V\gg V_c\,$. It is easy to see that the leading contribution
is of the form $V/V_c$ and is governed by Eq.~(\ref{Xi_2}).
Furthermore, only the term proportional to $|\Pi|^2$ should be
retained in $\Xi_z$, Eq.~(\ref{Xi_z}), to find the asymptotics.
Changing the variables $z_1$, $z_2$ to \mbox{$Z=(z_1+z_2)/\sqrt{2}\,$},
\mbox{$z=(z_1-z_2)/\sqrt{2}\,$}, using the expansion (\ref{Pi_z}),
and evaluating the integrals via the residue theorem, we get
\bea
\left\langle\delta g^2\right\rangle_{V\gg V_c} &=&
{}-32\,\frac{V}{V_c}\int\! dz\,\frac{\partial^2}{\partial\alpha^2}\,{\rm
  Tr}\,\left|\Pi_z\right|^2\Big|_{\alpha=0}\nonumber\\ &=&
c_1\,\frac{V}{V_c}\:\:,
\label{LK-coefficient}
\eea
where
\bea
c_1 &=&
64\sum\limits_{m,n>0}\frac{1}{\pi^5}
{\textstyle\left[\frac{1}{n^4(m^2+n^2)}-
\frac{1}{m^2n^2(m^2+n^2)}\right]}
\left\langle n|y|m\right\rangle^2\nonumber\\
&=& 7.785\cdot 10^{-4}\:\:. \label{c1}
\eea
Therefore, although the large-$V$ asymptotics is indeed of the form
$V/V_c$ obtained by LK, the corresponding numerical coefficient $c_1$ 
(not evaluated in Ref.~\onlinecite{Larkin_Khmelnitskii}) is very
small.

\begin{figure}[h]
\begin{center}
\includegraphics[width=1.0\linewidth]{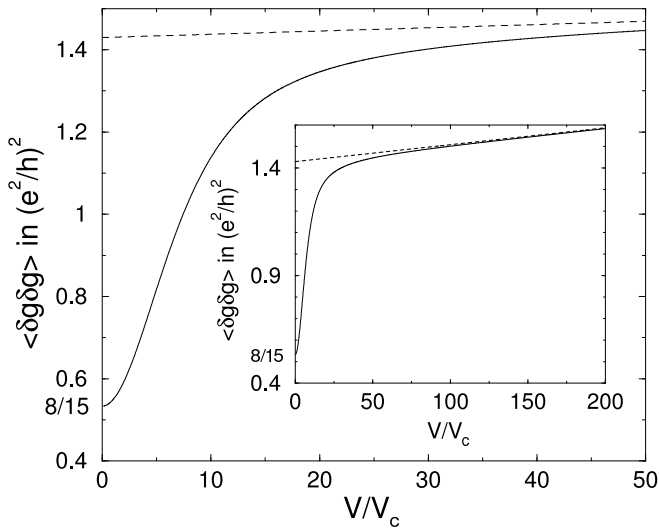}
\caption{\label{plot} Fluctuations $\langle\delta g^2\rangle$
of the differential conductance as
a function of the bias voltage normalized to the Thouless energy,
$V/V_c\,$.
The inset shows the emergence of the linear behavior at very large
$V/V_c\,$. The dashed line represents the asymptotic behavior,
\mbox{$\langle\delta g^2\rangle=8/15+c_0+c_1 V/V_c\,$}
with \mbox{$c_0=0.8964$} and \mbox{$c_1=7.785\cdot 10^{-4}$}.}
\end{center}
\end{figure}

We now turn to the evaluation of the full crossover behavior of
$\langle\delta g^2\rangle$ as a function of $V/V_c\,$. It is more
cumbersome, but can be done in a similar way; see
Appendix~\ref{app-crossover} for details.
The result is shown in Fig.~\ref{plot}.
The striking feature of this plot is that the large-$V$ asymptotics
(\ref{LK-coefficient}) becomes applicable only at very large values of
$V/V_c\,$. The emergence of the linear contribution with a coefficient
given by Eq.~(\ref{LK-coefficient}) can be seen for
\mbox{$V/V_c\gtrsim 50$} but it becomes dominant only at
\mbox{$V/V_c\gtrsim 1000\,$}. For \mbox{$V/V_c\gtrsim 50$} the
conductance fluctuations are well described by the asymptotic formula
which includes, in addition to the linear term (\ref{LK-coefficient}),
a $V$-independent contribution (which is formally of the next order
but numerically dominates up to  \mbox{$V/V_c\sim 1000$}),
\be
\left\langle\delta g^2\right\rangle_{V\gg V_c} = \frac{8}{15} + c_0 + c_1 \frac{V}{V_c}\:\:,
\label{c0c1}
\ee
with \mbox{$c_0=0.8964$}.
On the other hand, this very slow increase of
$\langle\delta g^2\rangle$ (barely distinguishable from a saturation)
at large voltages is preceded by a broad crossover regime at
\mbox{$V/V_c\lesssim 50\,$}, where the conductance increases by roughly
a factor of 3.
In the experiments, the increase of $\langle\delta g^2\rangle$ is
observed in the range \mbox{$V/V_c\lesssim 200\,$}.
As is obvious from Fig.~\ref{plot}, this enhancement should be mainly
attributed to the crossover regime rather than to the LK linear
asymptotics. We will return to the comparison with experiment later,
after having considered the effects of finite temperature and of the
interaction-induced dephasing.

\section{Finite temperature}
\label{finite-T}

In this Section, we address the effect of finite temperature on the
non-equilibrium conductance fluctuations. We assume that the
temperature is not too high, so that at low voltages the inelastic
scattering can be neglected (the dephasing length $L_\phi(T)$ is large
compared to the system size).
The temperature however influences the result via the change of the
electron distribution function in the reservoirs. 
At a finite temperature, Eqs.~(\ref{Xi_0})--(\ref{Xi_2}) are modified
in the following way (here we switch back to dimensionful voltages for
clarity),
\bea
\left\langle\delta g\,\delta
  g\right\rangle_0 &=& \frac{16}{e^2}\int d\epsilon_1d\epsilon_2\:\Delta
f'_1\,\Delta f'_2\:\Xi_{\epsilon_1-\epsilon_2}\:\:,\label{Xi_0_T}\\
\left\langle\delta g\,\delta g\right\rangle_1&=& \frac{16}{e^2}\int d\epsilon_1d\epsilon_2\,\Delta f'_1\Delta
f_2\,\frac{\partial}{\partial V_2}\Xi_{\epsilon_1-\epsilon_2}\nonumber\\
& &{}+\frac{16}{e^2}\int d\epsilon_1d\epsilon_2\,\Delta f_1\Delta
f'_2\,\frac{\partial}{\partial V_1}\Xi_{\epsilon_1-\epsilon_2},\label{Xi_1_T}\\
\left\langle\delta g\,\delta g\right\rangle_2 &=& \frac{16}{e^2}\int\!
d\epsilon_1d\epsilon_2\:\Delta f_1\Delta f_2\frac{\partial^2}{\partial
  V_1\partial V_2}\Xi_{\epsilon_1-\epsilon_2},\label{Xi_2_T}
\eea
where \mbox{$\Delta f_i=f(\epsilon_i)-f(\epsilon_i+eV_i)\,$}
($f$ is the Fermi function) and
\mbox{$\Delta f'_i=\frac{\partial}{\partial V_i}\Delta f_i\,$}.
Again the derivatives are taken at \mbox{$V_1=V_2=V\,$}.

Evaluation of the full crossover (i.e.~the conductance fluctuations at
arbitrary $V/V_c$ and $T/eV_c$) is too cumbersome. We thus restrict
ourselves to the limit \mbox{$T\gg eV_c\,$}, when the temperature
strongly affects the  conductance fluctuations at low voltages (in the
opposite limit, \mbox{$T\ll eV_c\,$}, the temperature is irrelevant
for all voltages, so that the results of Sec.~\ref{non-int} apply).
This condition is reasonably fulfilled in most of relevant
experiments, where the temperature is usually several times higher
than the Thouless energy.

We consider now the two limits of low \mbox{($V\ll V_c$)} and high
\mbox{($V\gg V_c$)} bias voltages.
In the first case, \mbox{$V/V_c \to 0\,$}, only the voltage-independent 
contribution $\left\langle\delta g\,\delta g\right\rangle_0$ survives.
For \mbox{$T\gg eV_c$} the thermal smearing strongly suppresses this
contribution compared to its zero-temperature (UCF) value,  
\be
\left\langle\delta g\,\delta g\right\rangle_0
\:\stackrel{T\gg eV_c}{=}\:\frac{8\pi}{9}\cdot\frac{eV_c}{T}\:\:.
\label{Xi_0_asymptotic}
\ee

In the opposite limit, when \mbox{$eV, T\gg eV_c$} (the ratio $eV/T$
can be arbitrary), we find that the high-voltage behavior (\ref{c0c1})
is modified by the temperature in the following way:
\be
\left\langle\delta g\,\delta g\right\rangle =
a\left(\frac{eV}{T}\right)\cdot c_0+
b\left(\frac{eV}{T}\right)\cdot c_1\frac{V}{V_c}\:\:,\:\:\:V\gg V_c
\label{high-T-high-V}
\ee
where the functions $a\left(\frac{eV}{T}\right)$ and
$b\left(\frac{eV}{T}\right)$ have the form
\bea
a\left(\frac{eV}{T}\right) &=& 2\int d\epsilon\:
\Big[f({\epsilon})-f(\epsilon+eV)\Big]\left(-\frac{\partial
  f}{\partial\epsilon}\right)\nonumber\\
 &=& \frac{{\rm sinh}\frac{eV}{T}-\frac{eV}{T}}{{\rm
    cosh}\frac{eV}{T}-1}\:\:,\label{a(V/T)}\\
b\left(\frac{eV}{T}\right) &=& \frac{1}{eV}\int
d\epsilon\:\big[f({\epsilon})-f(\epsilon+eV)\big]^2\nonumber\\
 &=& {\rm coth}\frac{eV}{2T}-\frac{2T}{eV}\:\:.\label{b(V/T)} 
\eea
It is worth emphasizing that for any $T$ at sufficiently high bias
voltages the $T=0$ result is recovered up to the missing offset of
$8/15\,$.

\begin{figure}[h]
\begin{center}
\includegraphics[width=1.0\linewidth]{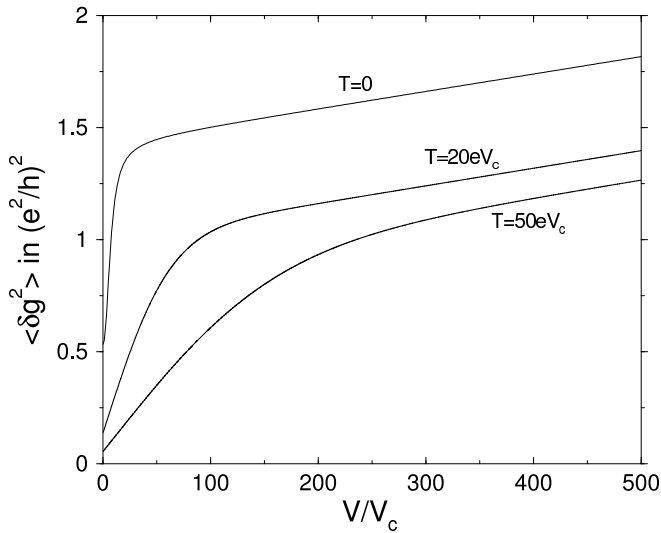}
\caption{\label{plot-T}
The voltage dependence of the conductance fluctuations as given by the
sum of Eqs.~(\ref{Xi_0_asymptotic}) and (\ref{high-T-high-V}),
for the temperatures $T=20\,{\rm eV_c}$ and $T=50\,{\rm eV_c}\,$.
The zero-T result of Fig.~\ref{plot} is also shown for comparison.
It is clearly seen that the coefficient of the asymptotic linear
behavior is not changed by the temperature.}
\end{center}
\end{figure}

Adding the term $\left\langle\delta g\,\delta g\right\rangle_0\,$,
Eq.~(\ref{Xi_0_asymptotic}), to (\ref{high-T-high-V}), we get an
interpolation formula which is parametrically justified in the regime
\mbox{$V\gg V_c$} and has the correct limit at \mbox{$V/V_c\to 0$}.
This formula can thus be used as a convenient approximation for
$\langle\delta g\,\delta g\rangle$ in the full range of voltages at
\mbox{$T\gg eV_c\,$}.
The resulting voltage dependence of the conductance fluctuations is
shown in Fig.~\ref{plot-T} for the temperatures $T=20\,{\rm eV_c}$ and 
$T=50\,{\rm eV_c}\,$.
We compare these theoretical results with experimental data in
Sec.~\ref{compare-exp} below.

\section{Dephasing by electron-electron scattering at high bias voltages}
\label{int}

Up to now, we have neglected the dephasing induced by the inelastic
scattering processes, which is justified for not too high bias voltages
$V$. With increasing $V$, the electron-electron scattering becomes
stronger, and eventually leads to suppression of the conductance
fluctuations. The corresponding theory is presented in this
Section. As in Sec.~\ref{finite-T}, we will assume that the
temperature at the reservoirs is sufficiently low (more precisely,
$T\ll geV_c$), so that the corresponding dephasing length is much
larger than the system size, $L_\phi(T)\gg L\,$. In this situation, the
dephasing will be entirely due to a high applied voltage. 

In quasi-one-dimensional diffusive systems, dephasing is dominated by
scattering processes with a {\em small} energy transfer
\cite{Altshuler_Aronov, AAK} \mbox{$\omega\ll T\,$},
which allows one to replace the dynamically screened Coulomb
interaction by a fluctuating classical potential.
The same arguments apply in the presence of a high bias voltage, when
the fluctuations of the electric field are characterized by the
effective temperature $T_{\rm eff}$ defined below.
In view of the semiclassical character of the problem, we can use the
Boltzmann-Langevin approach (see Refs.~\onlinecite{Kogan} and
\onlinecite{Nagaev}) to determine the correlation function of the
fluctuating field. The fluctuation of the local current density
$\mbox{\bf \em j}$ can be expressed in the following way,
\be
\delta \mbox{\bf \em j} = - i\mbox{\bf \em
q}D\,\delta\rho+\sigma\,\delta \mbox{\bf \em E}
+\delta \mbox{\bf \em j}^{\rm ext}\:,
\label{boltzmann-langevin}
\ee
where the first term is the diffusion flow produced by the
density fluctuation $\delta\rho$, the second one is due to the
fluctuation $\delta \mbox{\bf \em E}$ of electric field
($\sigma = e^2 \nu D$ is the conductivity), and the last
term represents extraneous (Langevin) current fluctuations induced by
elastic scattering. At equilibrium, these correlations are found from
the fluctuation-dissipation theorem.\cite{AAK} Out of equilibrium,
they have been studied in the shot noise context.
\cite{Blanter_Buttiker} We only need the fluctuations of
$x$-components of the current (those along the wire),
\be
\left\langle\delta j_{x}^{\rm ext}(x,t)\,\delta j_{x}^{\rm
    ext}(x',t')\right\rangle 
=2\sigma\,T_{\rm eff}\,\delta(x-x')\,\delta(t-t')\:,
\ee
with the effective temperature $T_{\rm eff}(x)$ given by
\be
T_{\rm eff}=\int d\epsilon\:n_\epsilon(x)\left[1-n_\epsilon(x)\right]\:,
\ee
where $n_\epsilon(x)$ is the (in general, nonequilibrium) 
distribution function. This function itself is strongly voltage
dependent.

For not too high $V$, the distribution function has a
double-step shape (\ref{double-step}); for higher voltages
the inelastic scattering processes lead to thermalization of
electrons, and the the distribution function assumes the Fermi shape
with a local temperature.\cite{Blanter_Buttiker} 
The boundary between the two regimes is determined by the energy
relaxation time $\tau_{\epsilon}$: no thermalization takes place for
\mbox{$eV_c\gg\tau_{\epsilon}^{-1}$}.
It is important that the dephasing is dominated by small energy
transfers \mbox{$\epsilon \ll T_{\rm eff}\,$}, which implies
\cite{AAK} that the dephasing time $\tau_{\phi}$ is much shorter than
the energy relaxation time $\tau_\epsilon$. Thus, even when the system
enters, with increasing voltage, the regime of strong dephasing,
\mbox{$eV_c \ll \tau_{\phi}^{-1}$}, the energy relaxation still
remains weak, \mbox{$eV_c\gg\tau_{\epsilon}^{-1}$}, in a
parametrically broad range of voltages.
From now on, we consider this regime, and find for the effective
temperature
\be
\label{teff}
T_{\rm eff}(x)=eV\frac{x}{L}\left(1-\frac{x}{L}\right)\:.
\ee
This effective temperature is position-dependent, taking its maximum
in the middle of the wire and approaching zero at the contacts with
the reservoirs. 
Using the continuity equation,
\be
\delta\rho=\frac{\mbox{\bf \em q}}{\omega}\,\delta\mbox{{\bf \em j}}\:,
\ee
and the relation between fluctuations of the electric field
\mbox{$\mbox{\bf \em E}=-{\bf\nabla}\phi$} and the density,
\be
e\,\delta\phi(\mbox{\bf \em q}) = 
\frac{\delta\rho({\mbox{\bf \em q}})}{e}\,U^{(0)}(q)
\ee
where \mbox{$U^{(0)}\left(|\mbox{\bf \em r--r}'|\right)
  =e^2/|\mbox{\bf \em r--r}'|$} is the bare Coulomb interaction, one
finds
\be
\delta j_{x}=\frac{-i\omega}{-i\omega+Dq^2\left[1+\nu
  U^{(0)}(q)\right]}\,\delta j_{x}^{\rm ext}\:.
\ee
Since for Coulomb interaction in metals \mbox{$\nu U^{(0)}(q)\gg 1\,$}, 
Eq.~(\ref{boltzmann-langevin}) reduces to
\be
\sigma\,\delta E_{x}(q)=-\delta j_{x}^{\rm ext}(q)\:,\qquad q\neq 0\:.
\ee
The $q=0\,$-component of \mbox{$\delta E_x=-\partial_x\phi$} is zero
due to the boundary conditions: there are no potential fluctuations at
the perfect reservoirs. Therefore, the electric field fluctuations in
the coordinate space are given by
\be
\delta E_{x}(x)=-\frac{1}{\sigma}\left[\,\delta j_{x}^{\rm
    ext}(x)-\frac{1}{L}\int dx\:\delta j_{x}^{\rm ext}(x)\,\right]\:\:.
\ee
Using Eq.~(\ref{teff}), we obtain the correlator of the electric fields,
\bea
\big\langle\delta E_{x}(x)\,\delta E_{x}(x')\big\rangle&=&
\frac{2}{\sigma}\bigg[T_{\rm eff}(x)\,\delta(x-x')-\frac{1}{L}T_{\rm
    eff}(x)\nonumber\\
  & &\!\!\!\!\!\!{}-\frac{1}{L}T_{\rm
    eff}(x')+\frac{1}{L^2}\!\!\int\!\!d\tilde{x}\,T_{\rm
    eff}(\tilde{x})\bigg]. 
\eea
The correlator of the potential can be found by spatial integration. We
introduce dimensionless variables \mbox{$y=x/L\,$}, \mbox{$\tau=eV_ct\,$},
\mbox{$\theta=\phi/eV_c\,$}, and find
\bea
\label{theta-corr}
\lefteqn{\big\langle\theta(y,\tau)\,\theta(y',\tau')\big\rangle
\:=\: \frac{2V}{3V_cg}\,\delta(\tau-\tau')\,y_<(1-y_>)}
\nonumber\\  
&&\qquad
\times\left[y(1-y)+y'(1-y')-\frac{\left(y_>-y_<\right)}{2}\right],
\eea
where \mbox{$y_<={\rm min}(y,y')$} and \mbox{$y_>={\rm max}(y,y')\,$}.

As can be seen from (\ref{theta-corr}), 
the dimensionless parameter $V/V_cg$ is a measure for the importance
of inelastic processes. Therefore, in the presence of the Coulomb
interaction, the results of Section~\ref{non-int} hold under the
condition \mbox{$V\lesssim V_cg\,$}. This condition excludes, in
particular, voltages of order $V\sim V_cg^2$ which are required to
reach the regime where the fluctuating differential resistance would
become negative in the absence of dephasing.

As we noted in the previous section, the part $\sim |\Pi|^2$ of
Eq.~(\ref{Xi_2}) 
dominates at high voltages, so that we can neglect
\mbox{${\rm Re}\,\Pi^2\,$}.
Following the approach of Refs.~\onlinecite{AAK,Aleiner_Blanter},
we express $\Pi$ as a path integral,
\begin{widetext}
\bea
{\rm Tr}\left|\Pi_z\right|^2 &=& \int\limits_0^1
dy_1\,dy_2\,\,\int\limits_0^\infty d\tau_1\,d\tau_2\:
 {\rm e}^{iz
   (\tau_1-\tau_2)}\int\limits_{\xi_1(0)=y_2}^{\xi_1(\tau_1)=y_1}
 {\cal D}\xi_1(t_1)\int\limits_{\xi_2(0)=y_2}^{\xi_2(\tau_2)=y_1}
 {\cal D}\xi_2(t_2)\:\times\nonumber\\
& &{\rm exp}\left\{\:\int\limits_0^{\tau_1}dt_1
\left[{}-\frac{\dot{\xi_1}^2}{4}-\frac{i(V_1-V_2)}{V_c}\,\xi_1(t_1)-i\theta(\xi_1(t_1),t_1)\right]\right.\nonumber\\
& &\left.\qquad{}+\int\limits_0^{\tau_2}dt_2 
\left[{}-\frac{\dot{\xi_2}^2}{4}+\frac{i(V_1-V_2)}{V_c}\,\xi_2(t_2)
  +i\theta(\xi_2(t_2),t_2)\right]\right\}\:\:\:.
\label{path_integral_nonav}
\eea
\end{widetext}
We now perform the averaging over the random fields by using
Eq.~(\ref{theta-corr}). When dephasing is strong, conductance
fluctuations are dominated by pairs of paths which stay close to each
other, $|\xi_1(t)-\xi_2(t)|\ll 1\,$. 
We thus expand the terms in the action representing the dephasing 
in  $|\xi_1(t)-\xi_2(t)|\,$, which yields
\begin{widetext}
\bea
\left\langle{\rm Tr}\left|\Pi_z\right|^2\right\rangle&=&\int\limits_0^1
dy_1\,dy_2\,\,\int\limits_0^\infty d\tau_1\,d\tau_2\,\,{\rm e}^{iz (\tau_1-\tau_2)}\int\limits_{\xi_1(0)=y_2}^{\xi_1(\tau_1)=y_1}{\cal
  D}\xi_1(t_1)\int\limits_{\xi_2(0)=y_2}^{\xi_2(\tau_2)=y_1}{\cal 
  D}\xi_2(t_2)\,\times\nonumber\\
& &{\rm
  exp}\left\{\:\int\limits_0^{\phantom{(}\tau_1\phantom{)}}dt_1
\left[{}-\frac{\dot{\xi_1}^2}{4}-
\frac{i(V_1-V_2)}{V_c}\,\xi_1(t_1)\right]\:\:
+\int\limits_0^{\phantom{(}\tau_2\phantom{)}}dt_2
\left[{}-\frac{\dot{\xi_2}^2}{4}+\frac{i(V_1-V_2)}{V_c}\,\xi_2(t_2)\right]\right.\nonumber\\
& &\qquad\left.{}-\frac{2V}{V_cg}\,F(y_2)
\big|\tau_1-\tau_2\big|\:\:\:-\frac{2V}{V_cg}\!\!\int\limits_0^{{\rm
    min}(\tau_1,\tau_2)}\!\!\!\!dt\:\big|\xi_1(t)-\xi_2(t)
\big|\:y_2(1-y_2)\right\}\:\:,
\label{path_integral}
\eea
\end{widetext}
\be
F(y)=\frac{2}{3}\,y^2\left(1-y\right)^2\:.
\ee
Note that the factors $F(y_2)$ and $y_2(1-y_2)$ in the dephasing
terms could be equally well written as  $F(y_1)$ and $y_1(1-y_1)$
respectively, since we assumed that the distance between $y_1$ and
$y_2$ is much smaller than their distance to the reservoirs,
$|y_1-y_2|\ll y_1,\,(1-y_1)\,$. We will have to check the consistency
of this assumption afterwards.   

The last two terms in the exponent of Eq.~(\ref{path_integral})
represent the dephasing effects. The first of them is the dephasing
during a small time difference $|\tau_1-\tau_2|$ between the time
extensions of the both diffuson trajectories, while the second one is
the dephasing during the motion of both diffusons.
Since after the $z$-integration the characteristic values of 
$|\tau_1-\tau_2|$ are of order $V_c/V$, the first term gives a
contribution of order $1/g$ to the exponent and thus can be neglected.

Employing the Fourier transform,
\be
\Xi_z=\int d\tau\:{\rm e}^{{\rm
    i}z\tau}\,\tilde{\Xi}(\tau)\:\:\:,\:\:\:
\int dz\:\Xi_z=2\pi\,\tilde{\Xi}(0)\:,
\ee
we can write the important part of Eq.~(\ref{Xi_2}) (the part
containing $|\Pi|^2$) for \mbox{$V\gg gV_c$} as
\be
\left\langle\delta g\,\delta g \right\rangle_2 =
32\pi\,VV_c\,\frac{\partial^2}{\partial V_1\partial V_2}
\tilde{\Xi}(0)\Big|_{V_1=V_2=V}\:, 
\ee
where 
$\tilde{\Xi}(0)$
is twice Eq.~(\ref{path_integral}) with $\tau_1$ set to $\tau_2\,$.

By changing to new variables \mbox{$\zeta_1=(\xi_1+\xi_2)/\sqrt{2}\,$},
\mbox{$\zeta_2=(\xi_1-\xi_2)/\sqrt{2}$} we can perform the path
integral over $\zeta_1\,$, which is not affected by the random fields
and just yields $\Theta(\tau)/\sqrt{2}\,$. Then
\be
\left\langle\delta g\,\delta g\right\rangle_2 =
32\sqrt{2}\,\pi VV_c\,\frac{\partial^2}{\partial V_1\partial V_2}
\int\limits_0^1 dy \int\limits_0^\infty d\tau\:
I_y(0,\tau)\Big|_{V_{1,2}=V}\:, 
\ee
where $I_y(\zeta,\tau)$ obeys the equation
\bea
\label{diff-eq}
\left[\partial_\tau-\partial_\zeta^2+i\sqrt{2}
{\textstyle\frac{(V_1-V_2)}{V_c}}\zeta+
{\textstyle\frac{2V}{V_cg}}\sqrt{2}\,y(1-y)\,|\zeta|\right]&&
\!\!\!\!\!I_y(\zeta,\tau)\nonumber\\
=\delta(\zeta)\,\delta(\tau)\:.& & 
\eea
This can be made dimensionless by choosing new variables
\mbox{$t=p^{2/3}\tau\,$}, \mbox{$\eta=p^{1/3}x\,$},
\mbox{${\cal I}=p^{-1/3}I\,$}, and
\mbox{$v=i\sqrt{2}(V_1-V_2)/V_cp\,$}, where
\mbox{$p=\sqrt{2}\,y(1-y)\,2V/V_cg\,$}.
Performing the time integration over $t$, we get
\be
\left\langle\delta g\,\delta g \right\rangle_2 =
8\pi\left(\frac{V_c}{V}\right)^{4/3}\!g^{7/3}
\frac{\partial^2}{\partial v^2}Q(0)\Big|_{v=0}
\int\limits_0^1\!\!\!\frac{dy}{\left[y(1-y)\right]^{7/3}},
\ee
where $Q$ satisfies
\mbox{$(-\partial_\eta^2+|\eta|+v\eta)\,Q(\eta)=\delta(\eta)\,$}.
We see that the resulting integral diverges at the upper and lower
limits corresponding to the vicinity of the reservoirs. 
The reason for this divergence is as follows:
Near the reservoirs the expansion of the action in
\mbox{$|\xi_1(t)-\xi_2(t)|\,$}, used to derive
Eq.~(\ref{path_integral}), ceases to be valid, since the
characteristic values of \mbox{$|\xi_1(t)-\xi_2(t)|$} determined by
the local value of the dephasing length $L_\phi(y)$ become of order of
$y$ or $(1-y)\,$.
The above divergence shows that these small vicinities of reservoirs
(where the effective temperature approaches zero and the dephasing
length is larger than in the rest of the wire) in fact dominate the
conductance fluctuations.

The exact calculation thus requires using the full form of the
correlation function (\ref{theta-corr}), which makes the problem much
more cumbersome, leading  to a two-dimensional Schr\"odinger-type
differential equation rather than to the one-dimensional equation
(\ref{diff-eq}).
We can get, however, the result, up to a numerical prefactor of order
unity, in a simpler way.
Indeed, as follows from the above consideration, the integral must be
cut off at a distance $Ly_c$ to the reservoirs which is of the order
of the local phase-breaking length, yielding the self-consistency
condition
\be
Ly_c \sim L_\phi(y_c) = \sqrt{D\tau_\phi[T_{\rm eff}(y_c)]}\:\:,
\ee
where $T_{\rm eff}(y)=eVy(1-y)$ and
\mbox{$\tau_\phi[T_{\rm eff}]\sim\left(D\nu^2/T_{\rm eff}^2\right)^{1/3}$}.
This yields a cutoff
\be
\label{yc}
y_c\sim\left(\frac{V_cg}{V}\right)^{1/4}\:.
\ee
Therefore, the amplitude of conductance fluctuations, which are
dominated by the coherent segments within the distance $Ly_c$ to the
contacts, is given by
\be
\left\langle\delta g\,\delta g\right\rangle_2 \:\sim\:
g^2\,\frac{V_c}{V}\:\:,\qquad \frac{V}{V_c} \gg g\:.
\label{first_term}
\ee
Thus, the variance of the conductance shows a maximum at
\mbox{$V/V_c\sim g\,$} and then decays according to the $1/V$ power
law. 

Strictly speaking, in order to show rigorously that the asymptotic
suppression at high bias voltages is given by this power law, we still
have to check that
\mbox{$\left\langle\delta g\,\delta g\right\rangle_1$} and
\mbox{$\left\langle\delta g\,\delta g\right\rangle_0$}
remain smaller than \mbox{$\left\langle\delta g\,\delta g\right\rangle_2$}
with increasing voltage. This turns out to be the case, as is
shown in Appendix~\ref{app-subleading_terms}.

A natural question to be asked at this point is whether the result 
(\ref{first_term}) can be obtained from simple qualitative
arguments. Indeed, we know that for the case of the linear-response
transport, when the dephasing length $L_\phi$ is governed by the
temperature $L_\phi=L_\phi(T)\,$, the conductance fluctuations of a
wire in the high-temperature regime, $L_\phi(T)\ll L\,$, can be
estimated in the following way. The system is split into segments of
the length $L_\phi\,$. The electrons within each of the segments remain
phase coherent, yielding fluctuations of the conductance of the
segment, $\langle \delta g^2(L_\phi)\rangle\sim D/L_\phi^2T$, see 
Eq.~(\ref{Xi_0_asymptotic}). A fluctuation $\delta g(L_\phi)$ of the
conductance of a segment induces a fluctuation in the total
conductance 
\be
\label{qualit0}
\delta g \sim \left(L_\phi/L\right)^2 \delta g(L_\phi)\:.
\ee
 Adding
incoherently fluctuations of all $L/L_\phi$ segments, one gets
\be
\label{qualit1}
\langle \delta g^2\rangle \sim \left(L_\phi/L\right)^3
\langle \delta g^2(L_\phi)\rangle
\sim \frac{D}{L^2 T}\frac{L_\phi}{L}\:.
\ee
This agrees with the diagrammatic calculation
\cite{Lee_Stone_Fukuyama} (using $L_\phi$ as a phenomenological
parameter), as well as with the path integral 
calculation,\cite{Aleiner_Blanter} yielding
the dephasing length $L_\phi(T)$ induced by the electron-electron
scattering, $L_\phi(T)\sim(D^2\nu/T)^{1/3}$. 

A generalization of this consideration on the strongly non-equilibrium
setup was attempted in Ref.~\onlinecite{Strunk} but the obtained
result (mentioned in Sec.~\ref{intro}) was different from
Eq.~(\ref{first_term}). The arguments presented in
Ref.~\onlinecite{Strunk} miss the following two non-trivial properties
of the non-equilibrium dephasing, which distinguish it from its
equilibrium, finite-$T$ counterpart.
First, as we have shown above, the dephasing length in non-equilibrium is
strongly position-dependent, and the conductance fluctuations are
dominated by the segments of the wire adjacent to the reservoirs.
Therefore, representing the wire as an incoherent series
of $L/L_\phi(eV)$ equal segments of a length $L_\phi(eV)$ is not
justified. Second, the conductance fluctuations in a segment of a
length $L_\phi$ cannot be obtained simply using the results for a
coherent sample of a length $L_\phi$ subject to a voltage
$V(L_\phi)=VL_\phi/L\,$. This is because the non-equilibrium
distribution function in the segment $L_\phi$ is {\it not} determined
by a single parameter $V(L_\phi)$ but rather is governed by two
parameters, $V$ and $L_\phi/L\,$. 

We are now going to show that when these peculiarities are properly
taken into account, the qualitative consideration does reproduce the
result (\ref{first_term}). Consider a coherent segment of a length
$L_\phi$ (to be specified later). Generalizing the consideration in
Sec.~\ref{finite-T}, the variance of its differential conductance can
be estimated as  
\be
\label{qualit2}
\langle\delta g^2(L_\phi)\rangle\sim 
\frac{1}{eV_c(L_\phi)}\int d\epsilon\,\left[\Delta f_{L_\phi}(\epsilon)\right]^2\:,
\ee
where $ eV_c(L_\phi)=D/L_\phi^2$ is the Thouless energy corresponding
to the segment $L_\phi$, and $\Delta f_{L_\phi}(\epsilon)$ is the
difference between the distribution function at the left and right
boundaries of this segment. It is easy to see that 
$\Delta f_{L_\phi}(\epsilon)$ is equal to $L_\phi/L$ in the energy
window of the width $eV$, and is zero outside this window. This yields
\be
\label{qualit3}
\langle\delta g^2(L_\phi)\rangle\sim \frac{1}{eV_c(L_\phi)}\,eV 
\left(\frac{L_\phi}{L}\right)^2\sim \frac{V}{V_c}\left(\frac{L_\phi}{L}\right)^4. 
\ee
According to Eq.~(\ref{qualit0}), this induces fluctuations of the
wire conductance of the magnitude
\be
\label{qualit4}
\langle\delta g^2\rangle \sim \left(\frac{L_\phi}{L}\right)^8 
\frac{V}{V_c}\:.  
\ee
Finally, using that $L_\phi=Ly_c$ with $y_c$ given by Eq.~(\ref{yc}),
we get $\langle\delta g^2\rangle \sim g^2V_c/V\,$, thus reproducing
the result (\ref{first_term}) of the path-integral calculation. 

\section{Comparison with experiment}
\label{compare-exp}

In this Section we compare our results with the measurements presented
in Refs.~\onlinecite{Strunk} and \onlinecite{Terrier_PhD}. There,
magnetoconductanc traces of a gold wire with a length $L=1.5\,\mu{\rm m}$
and a Thouless energy
of $eV_c\equiv D/L^2=3.4\,{\rm \mu eV}$ 
were taken at \mbox{$T=300\,{\rm mK}$}
over a voltage range up to \mbox{$V\simeq 3.7\,{\rm mV}$}. This yields
\mbox{$T/eV_c\simeq 7.6$} and the maximum value of
\mbox{$V/V_c\approx 1000\,$}. The dimensionless conductance of the
sample was \mbox{$g\approx 1400\,$}.

\begin{figure}[h]
\begin{center}
\includegraphics[width=1.0\linewidth]{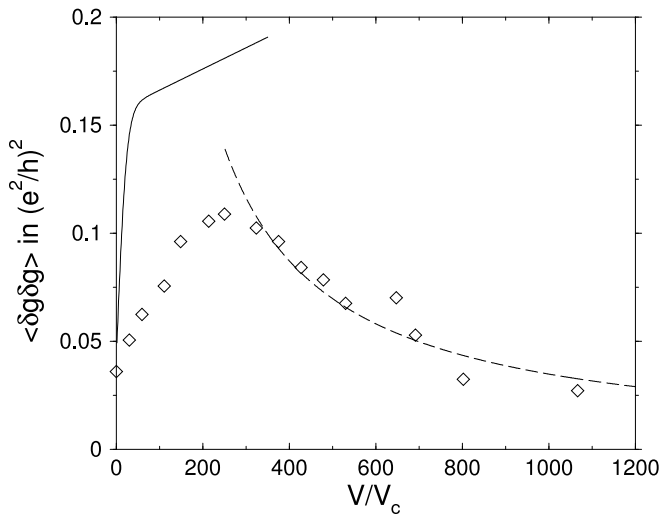}
\caption{\label{plot-compare} Comparison of our results with
  experimental data taken from Refs.~\onlinecite{Strunk,Terrier_PhD}.
  The diamonds are the data from Fig.~5.4 of
  Ref.~\onlinecite{Terrier_PhD}.
  The solid line represents the interpolation formula obtained by
  adding Eqs.~(\ref{Xi_0_asymptotic}) and (\ref{high-T-high-V}),
  multiplied by $1/8$ to account for spin-orbit interaction
  and broken time-reversal symmetry.
  The dashed line represents the asymptotic suppression,
  Eq.~(\ref{first_term}),
  with the numerical coefficient chosen from the fit to the data.}
\end{center}
\end{figure}

Fig.~\ref{plot-compare} shows our results for both the non-interacting
limit (not too high voltages) and the strong-dephasing limit (very
high voltages), compared with the measured data.
It is seen that there is a reasonably good agreement between the theory
and the experiment in the overall shape of $\langle\delta g\,\delta
g\rangle (V/V_c)$ and in the magnitude of its enhancement at the
maximum. 

The main deviation is that the initial increase of the conductance
variance is considerably less steep than on the theoretical
curve. This discrepancy can be partly attributed to the following
factors.
First, we used for the full curve in Fig.~\ref{plot-compare} the
formula (\ref{high-T-high-V}) which takes into account the crossover
at \mbox{$eV\sim T$} but not the crossover at \mbox{$V\sim V_c\,$}.
Though the temperature was larger than $eV_c$ by a factor $\sim 8$ 
in the experiment, we know from Sec.~\ref{non-int} that the $V/V_c$
crossover region is rather broad, extending up to
\mbox{$V/V_c\sim 50\,$}.
Thus the $V/V_c$ crossover is expected to overlap with the $eV/T$
crossover under the experimental conditions, making the total
crossover region broader.
Second, we did not take into account the effect of dephasing in the
low-voltage region, where we used the result for noninteracting
electrons. It is easy to see from Eq.~(\ref{path_integral}) that this
will lead to a relative correction of order $V/V_cg\,$, which will
provide a smooth matching with the strong-dephasing (high voltage)
regime (\ref{first_term}).

We also note that the linear behavior (\ref{LK-coefficient}) 
could not be accessed reliably in this experiment, since it requires,
in view of the very small value of the numerical coefficient
(\ref{c1}), very large voltages \mbox{$V/V_c\gtrsim 1000\,$}, while
the dephasing was setting in at voltages several times smaller. A
reliable observation of this linear behavior predicted by LK thus
requires samples with still larger values of the dimensionless
conductance $g$.

At high voltages, the experimentally observed decay of the conductance
fluctuations is consistent with our prediction of the $1/V$ decay, 
Eq.~(\ref{first_term}), see Fig.~\ref{plot-compare}. Indeed, the value
$\gamma=1.28\pm 0.12$ of the exponent of the power-law decay, $\langle
\delta g^2\rangle\propto V^{-\gamma}$ is in good agreement with our
result, $\gamma=1$. We thus disagree with the conclusion of the
authors of Ref.~\onlinecite{Strunk} who argued, based on a naive
estimate of the dephasing effect, that the electron-electron
scattering is not sufficient to convert the LK enhancement mechanism
into the observed fall-off, and ascribed this behavior to
electron-phonon scattering.
Experiments on samples with lower conductances might give more data in
the strong-dephasing range and thus allow a more accurate experimental
determination of the exponent of the power law. 

\section{Summary}
\label{summary}

In this article, we investigated 
the amplitude of mesoscopic fluctuations of the differential
conductance of a metallic wire at arbitrary bias voltage $V$. 
For non-interacting electrons, we have found, by using the Keldysh
diagram technique, that
the variance $\langle\delta g^2\rangle$
increases monotonously with $V$. The asymptotic large-$V$ behavior is 
\mbox{$\langle \delta g^2\rangle \sim V/V_c$} (where \mbox{$eV_c=D/L^2$}
is the Thouless energy), in agreement with the prediction by Larkin and
Khmelnitskii. We find, however, that this asymptotics has a very small
numerical prefactor and sets in at very large $V/V_c$ only, so that
it is very difficult to observe it reliably in the experiment. 
This high-voltage behavior is
preceded by a crossover regime, \mbox{$V/V_c\lesssim 30\,$}, where the
conductance variance increases by a factor $\sim 3$ as compared to
its value in the regime of universal conductance fluctuations (i.e, at
\mbox{$V\to 0$}). We further analyze, in the framework of the
path-integral technique,  the effect of dephasing due to the
electron-electron interaction on $\langle\delta g^2\rangle$ at high
voltages. With the Coulomb interaction taken into account, the
amplitude of conductance fluctuations becomes a nonmonotonic
function of $V$. Specifically, $\langle\delta g^2\rangle$ shows a
maximum at \mbox{$V/V_c\sim g$} and drops as $1/V$ for higher
voltages. In this regime, the conductance fluctuations are
dominated by quantum-coherent regions of the wire of a length
\mbox{$\sim L\,(gV_c/V)^{1/4}$} adjacent to the reservoirs.
Our results are in good agreement with available experimental data. 

\begin{acknowledgments}

Useful discussions with I.V.~Aleiner, B.L.~Altshuler, C.~Strunk, and
C.~Sch\"onenberger are gratefully acknowledged.
We also thank C.~Strunk and C.~Sch\"onenberger for providing us with
details of experimental data of Refs.~\onlinecite{Strunk,Terrier_PhD}.
Part of this work was done when the authors participated at the
Workshop ``Quantum Transport and Correlations in Mesoscopic Systems
and Quantum Hall Effect''
at the Max-Planck-Institut f\"ur Physik Komplexer Systeme in Dresden. 

\end{acknowledgments}

\appendix

\section{Keldysh diagram for the current}
\label{app-current}

Here we give the derivation of Eq.~(\ref{e1}), which relates the
current to the impurity averaged diagonal Keldysh function.

\begin{figure}[h]
\begin{center}
\includegraphics[width=1.0\linewidth]{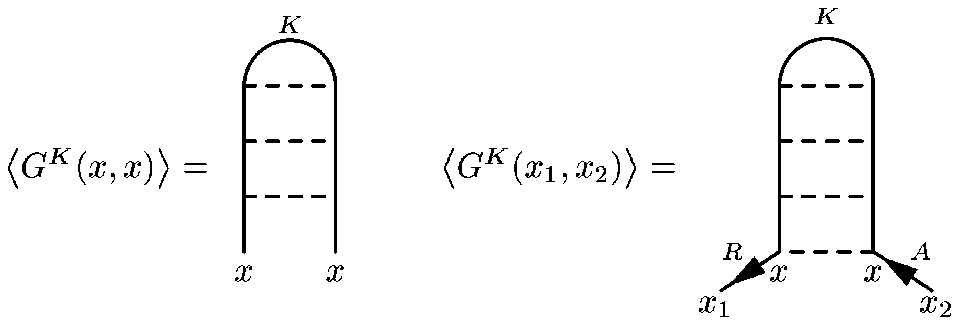}
\caption{\label{diag_K}
Diagrammatic representation of the relation between the averaged
Keldysh function $\left\langle G^K(x_1,x_2)\right\rangle$ and the
averaged diagonal Keldysh function
$\left\langle G^K(x,x)\right\rangle\,$,
Eq.~(\ref{diagonal_Keldysh}), which is used in the derivation of
Eq.~(\ref{e1}).}
\end{center}
\end{figure}

We begin by expressing the averaged Keldysh function
$\left\langle G^K(x_1,x_2)\right\rangle$ using the averaged diagonal Keldysh
function $\left\langle G^K(x,x)\right\rangle$.
The off-diagonal components of the kinetic equations for the
matrix Green's function add up to 
\be
(\epsilon-\hat{H})G^K+G^K(\epsilon-\hat{H})=
\frac{i}{\tau}\!\left[(2n_\epsilon-1)G^A+G^R(2n_\epsilon-1)\right]
\ee
(here impurity averaging is implied).
Using the gradient expansion,\cite{Rammer_Smith}
one finds
\bea
G^K(x,p)\!\!&=&\!\!\frac{i}{\tau}\Big\{\left[2n_\epsilon(x)-1\right]G^R(p)G^A(p)\nonumber\\
&+&\!\!\!i\partial_x n_\epsilon(x)\left[G^R(p)\partial_p
    G^A(p)-G^A(p)\partial_p G^R(p)\right]\!\!\Big\}.\nonumber\\
& &
\eea
This is equivalent to the relation
\bea
\lefteqn{\left\langle G^K(x_1,x_2)\right\rangle=}\nonumber\\
& &\!\!\!\!\!\!\!\!\!\!\frac{1}{2\pi\nu\tau}\!\int\! dx
\left\langle G^R(x_1,x)\right\rangle
\left\langle G^K(x,x)\right\rangle
\left\langle G^A(x,x_2)\right\rangle\,.
\label{diagonal_Keldysh}
\eea
This relation can be easily understood diagramatically, see
Fig.~\ref{diag_K}.
Using the relation\cite{Rammer_Smith} between the function
\mbox{$G^<\equiv (G^K+G^A-G^R)/2$} and the electron density matrix $\rho$,
\be
G^<(x_1,t;x_2,t)=iN\rho(x_1,x_2,t)
\ee
($N$ is the particle number), one finds
\be
I_x(x)=\frac{(-e)}{m}\int\frac{d\epsilon}{2\pi}\:
\frac{\partial}{\partial x}G^<(x,x')\Big|_{x'=x}\:.
\ee
The contribution from the term \mbox{$G^A-G^R$} is proportional to the
spectral density and does not depend on the state of the system.
Therefore only $G^K$ contributes to the current,\cite{Rammer_Smith}
\be
\left\langle I_x(x)\right\rangle=
\frac{(-e)}{2m}\,\frac{\partial}{\partial x}\int\frac{d\epsilon}{2\pi}
\left\langle G^K(x,x')\right\rangle\Big|_{x'=x}\:.
\label{av_current}
\ee
Inserting Eq.~(\ref{diagonal_Keldysh}) into Eq.~(\ref{av_current})
gives Eq.~(\ref{e1}).

\section{Evaluation of the Keldysh diagrams}
\label{app-diagrams}

In this Appendix we calculate the diagrams which arise in the Keldysh
technique and contribute to the correlation function of currents
(see Sec.~\ref{non-int}). 
We will show that the sum of all the diagrams yields Eq.~(\ref{II}).

We begin by presenting the expressions for the vertex factors (Hikami
boxes). Because of their local character (the electron Green's
function decays exponentially on the scale of the mean free path $l$,
which is much smaller than the system size $L$), they can be 
calculated in the momentum space.

The right Hikami box (containing two Keldysh vertices) 
of the diagram a-diff is of second order in the momenta:
\vspace*{1mm}

\parbox{2.5cm}{
\includegraphics[width=1.0\linewidth]{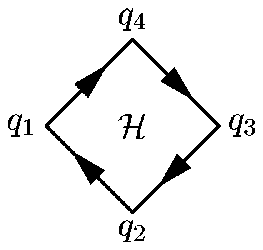}
}
$\qquad = 2\pi\nu D\tau^4\left(-2q_2q_4+q_1^2+q_3^2\right)\:$,\\
where the $q_i$ are the incoming momenta and \mbox{$q_1+q_2+q_3+q_4=0\,$}.
The left box (containing two velocity vertices) is of zeroth order: 
\vspace*{2mm}

\parbox{2.5cm}{
\includegraphics[width=1.0\linewidth]{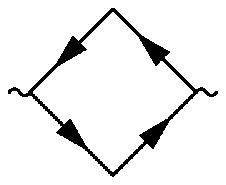}
}
$\qquad = 4\pi\nu D\tau^2\:$.
\vspace*{2mm}

Here $D$ is the diffusion constant, $\nu$ is the density of states and
$\tau^{-1}$ is the scattering rate characterizing the disorder
strength.
For the diagram b-diff the vertex factors are of first order in the
momenta:
\vspace*{1mm}

\parbox{2.5cm}{
\includegraphics[width=0.9\linewidth]{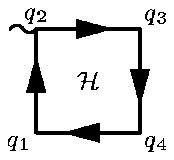}
}
$\qquad = 4\pi i\nu D\tau^3 q_4\:$.
\vspace*{1mm}

The vertex factors of a-coop and b-coop are the same as the vertex
factors of b-diff up to extra signs arising from the direction of the
Greens functions.

The vertex factors of the diagrams c-diff and c-coop are only needed
to zeroth order. Since these diagrams contain only one
diffuson/Cooperon, this gives a result of the same order in the
momenta as the other diagrams (containing one more diffuson/Cooperon)
evaluated up to second order in the momenta.
\vspace*{1mm}

\parbox{2cm}{
\includegraphics[width=0.9\linewidth]{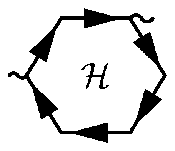}
}
$\qquad = 0\:$,
\vspace*{1mm}

\parbox{2cm}{
\includegraphics[width=1.0\linewidth]{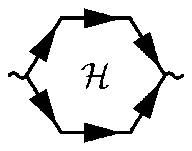}
}
$\qquad = -4\pi\nu D\tau^4\:$.
\vspace*{1mm}

Using these vertex factors and changing to real space representation,
the expressions corresponding to the diagrams are
\begin{widetext}
\bea
\left\langle\delta I\,\delta I\right\rangle_{\rm a-diff} &=&
\left(\frac{e}{4\pi 
    i}\right)^2\:\left(\frac{1}{2\pi\nu\tau}\right)^2\:4\pi\nu
D\tau^4\:2\pi\nu D\tau^2\int d\epsilon_1d\epsilon_2 \int
\frac{dx_1dx_2}{L^2}\times\nonumber\\ 
& &\qquad\bigg[\:2\,\frac{\partial}{\partial
  x_1}G^K_{\epsilon_1}(x_1)\:\frac{\partial}{\partial
  x_2}G^K_{\epsilon_2}(x_2)\:
{\cal P}_{\epsilon_1-\epsilon_2}\:{\cal P}_{\epsilon_2-\epsilon_1}\nonumber\\  
& &\qquad\:\:{}-\frac{\partial^2}{\partial
    x_1^2}{\cal P}_{\epsilon_1-\epsilon_2}\:
{\cal P}_{\epsilon_2-\epsilon_1}\,G^K _{\epsilon_1}(x_1)\:G^K_{\epsilon_2}(x_2)\:
-{\cal P}_{\epsilon_1-\epsilon_2}\:\frac{\partial^2}{\partial
    x_2^2}{\cal P}_{\epsilon_2-\epsilon_1}\:
G^K_{\epsilon_1}(x_1)\:G^K_{\epsilon_2}(x_2)\,\bigg]\:\:,\\
\left\langle\delta I\,\delta I\right\rangle_{\rm b-diff} &=&
\left(\frac{e}{4\pi i}\right)^2\:\left(\frac{1}{2\pi\nu\tau}\right)^2\:
\left(4\pi\nu D\tau^3\right)^2\int d\epsilon_1d\epsilon_2 \int\frac{dx_1dx_2}{L^2}\times\nonumber\\
& & \qquad\left[\frac{\partial}{\partial x_1}{\cal
    P}_{\epsilon_1-\epsilon_2}\:\frac{\partial}{\partial x_2}{\cal
    P}_{\epsilon_1-\epsilon_2}\:G^K_{\epsilon_1}(x_1)\:G^K_{\epsilon_2}(x_2)\:\:+\:{\rm c.c.}\:\right]\:\:,\\
\left\langle\delta I\,\delta I\right\rangle_{\rm c-diff} &=& 0\:\:,
\eea
\bea
\left\langle\delta I\,\delta I\right\rangle_{\rm a-coop} &=& \left(\frac{e}{4\pi
    i}\right)^2\:\left(\frac{1}{2\pi\nu\tau}\right)^2\:\left(4\pi\nu D\tau^3\right)^2\int\!d\epsilon_1d\epsilon_2 \int\!\frac{dx_1dx_2}{L^2}\:
{\cal P}_{\epsilon_1-\epsilon_2}\:{\cal P}_{\epsilon_2-\epsilon_1}\:\frac{\partial}{\partial x_1}G^K_{\epsilon_1}(x_1)\:\frac{\partial}{\partial x_2}G^K_{\epsilon_2}(x_2)\,,\\
\left\langle\delta I\,\delta I\right\rangle_{\rm b-coop} &=& \left(\frac{e}{4\pi
    i}\right)^2\left(\frac{1}{2\pi\nu\tau}\right)^2\left(4\pi\nu D\tau^3\right)^2\!\int\!d\epsilon_1d\epsilon_2\!\int\!\frac{dx_1dx_2}{L^2}
\left(\frac{\partial^2}{\partial x_1 \partial x_2}{\cal P}_{\epsilon_1-\epsilon_2}\right){\cal P}_{\epsilon_1-\epsilon_2}\,G^K_{\epsilon_1}(x_1)\,G^K_{\epsilon_2}(x_2),\\
\left\langle\delta I\,\delta I\right\rangle_{\rm c-coop} &=& \left(\frac{e}{4\pi
    i}\right)^2\:\left(\frac{1}{2\pi\nu\tau}\right)^2\left(-4\pi\nu D\tau^4\right)\int d\epsilon_1d\epsilon_2 \int \frac{dx_1dx_2}{L^2}\:
\delta(x_1-x_2)\:{\cal P}_{\epsilon_1-\epsilon_2}\:G^K_{\epsilon_1}(x_1)\:G^K_{\epsilon_2}(x_2)\:,
\eea
where ${\cal P}_{\epsilon_1-\epsilon_2}(x_1,x_2)$ is the diffusion
propagator satisfying the equation 
\be
\left\{D\frac{\partial^2}{\partial x_1^2}+i\Big[\epsilon_1+\phi_1(x_1)-\epsilon_2-\phi_2(x_1)\Big]\right\}\,{\cal P}_{\epsilon_1-\epsilon_2}(x_1,x_2) = -\left(2\pi\nu\tau^2\right)^{-1}\delta(x_1-x_2)\:\:,
\ee
and $\phi_{1,2}$ are the electrostatic potentials corresponding to the
voltages $V_{1,2}\,$.
Using the identity
\be
\frac{1}{4}\,\frac{\partial^2}{\partial x_1\partial x_2}{\cal P}^2(x_1,x_2)
= \left[\,\frac{\partial^2}{\partial x_1\partial x_2}{\cal P}(x_1,x_2)
  -\frac{1}{4\pi\nu D\tau^2}\,\delta(x_1-x_2)\,\right]{\cal P}(x_1,x_2)\:\:,
\label{identity}
\ee
we obtain
\bea
\left\langle\delta I\,\delta I\right\rangle_{\rm a-diff\,+\,b-diff}
&=& -\left(D\tau^2\right)^2\int d\epsilon_1
d\epsilon_2\left[\left|{\cal P}_{\epsilon_1-\epsilon_2}\right|^2 + \frac{1}{2}\,{\rm
  Re}\,{\cal P}^2_{\epsilon_1-\epsilon_2}\right]\frac{\partial}{\partial
  x_1}G^K(x_1)\:\frac{\partial}{\partial x_2}G^K(x_2)\:\:,\\
\left\langle\delta I\,\delta I\right\rangle_{\rm c-diff} &=& 0\:\:,\\
\left\langle\delta I\,\delta I\right\rangle_{\rm a-coop} &=&
-\left(D\tau^2\right)^2\int d\epsilon_1
d\epsilon_2\int\frac{dx_1dx_2}{L^2}\:\left|{\cal
    P}_{\epsilon_1-\epsilon_2}\right|^2\,
\frac{\partial}{\partial x_1}G^K(x_1)\:\frac{\partial}{\partial x_2}G^K(x_2)\:\:,\\
\left\langle\delta I\,\delta I\right\rangle_{\rm b-coop\,+\,c-coop} &=& -\left(D\tau^2\right)^2\int d\epsilon_1 d\epsilon_2\int\frac{dx_1dx_2}{L^2}\:\frac{1}{2}\,{\rm Re}\,{\cal P}^2_{\epsilon_1-\epsilon_2}\:\frac{\partial}{\partial x_1}G^K(x_1)\:\frac{\partial}{\partial x_2}G^K(x_2)\:\:.
\eea
\end{widetext}
Adding up these equations and rescaling the propagator by a factor
$2\pi\nu D\tau^2\,$, \mbox{$\Pi=2\pi\nu D\tau^2{\cal P}\,$}, we obtain
Eq.~(\ref{II}), which is written in units of
$\left(e^2/(2\pi)\right)^2=\left(e^2/h\right)^2$.

\section{Crossover from the linear response to the
  Larkin-Khmelnitskii regime}
\label{app-crossover}

In this Appendix we calculate the variance $\langle\delta g^2\rangle$ 
as a function of the bias voltage in the full range from the UCF
regime \mbox{($V/V_c\ll 1$)} to the LK asymptotic regime \mbox{($V/V_c\gg 1$)}.
In addition to the asymptotics Eq.~(\ref{LK-coefficient}), there are
contributions to the variance of the conductance which do not grow
asymptotically as $V/V_c\,$ but dominate in the intermediate regime.
First of all, there is a constant contribution 
\be
\label{ucf}
\left\langle\delta g\,\delta
g\right\rangle_0=\frac{8}{15}\:\:,
\ee 
which gives the familiar UCF result in the limit of
zero bias voltage. 
Second, there is a contribution from the term
$\left\langle\delta g\,\delta g\right\rangle_1$
containing one energy integration, which can be evaluated as
\begin{widetext}
\bea
\left\langle\delta g\,\delta g\right\rangle_1&=&32\int\limits_0^{V/V_c}
\!\!dz\:\frac{\partial}{\partial\alpha}\,{\rm
  Tr}\left[\,2\left|\Pi_{\frac{V}{V_c}-z}\right|^2+{\rm
  Re}\,\Pi_{\frac{V}{V_c}-z}^2\right]_{\alpha=0}\nonumber\\
&=&\frac{720+16\left(\frac{V}{V_c}\right)^2-360\left(-1\right)^{1/4}\sqrt{\frac{V}{V_c}}\left\{{\rm
      cot}\left[\left(-1\right)^{1/4}\sqrt{\frac{V}{V_c}}\,\right]+{\rm
      coth}\left[\left(-1\right)^{1/4}\sqrt{\frac{V}{V_c}}\,\right]\right\}}{45\left(\frac{V}{V_c}\right)^2}\label{1_Abs}\\
& &{}+\frac{8}{45}\:-16\sum\limits_{n=1}^\infty\frac{\left(\frac{V}{V_c}\right)^2-n^4\,\pi^4}{\left[n^4\,\pi^4+\left(\frac{V}{V_c}\right)^2\right]^2}\:\:\:,\label{1_Re}
\eea
\end{widetext}
where Eqs.~(\ref{1_Abs}) and (\ref{1_Re}) arise from the
contributions of $|\Pi|^2$ and ${\rm Re}\,\Pi^2\,$, respectively.
In the limit $V/V_c\to\infty\,$, the sum of Eqs.~(\ref{1_Abs}) and 
(\ref{1_Re}) saturates at $8/15\,$.
Then there is the part of $\left\langle\delta g\,\delta
g\right\rangle_2$ containing ${\rm Re}\,\Pi^2\,$, which also does not
contribute to the LK asymptotic behavior but gives a contribution
which saturates towards a constant as the voltage is increased:
\begin{widetext}
\bea
\left\langle\delta g\,\delta g\right\rangle_{2,\:{\rm
    Re}}&=&{}-16\int\limits_0^{V_1/V_c}\!\!dz_1
\int\limits_0^{V_2/V_c}\!\!dz_2\:\frac{\partial^2}{\partial\alpha^2}\,{\rm Tr}\:{\rm
    Re}\:\Pi_{z_1-z_2}^2\Big|_{\alpha=0}\nonumber\\
&=&32\sum\limits_{m,n>0}\frac{2\left(\frac{V}{V_c}\right)^2\left[m^4\pi^4+m^2n^2\pi^4+n^4\pi^4+\left(\frac{V}{V_c}\right)^2\right]}{m^2\,n^2\,\pi^4\left[m^4\pi^4+\left(\frac{V}{V_c}\right)^2\right]\left[n^4\pi^4+\left(\frac{V}{V_c}\right)^2\right]}\:\Big(\langle n|y|m\rangle\Big)^2\nonumber\\
&\stackrel{V\to\infty}{\longrightarrow}& 0.1905\:\:,
\label{2_Re}
\eea
\end{widetext}
where again \mbox{$|n\rangle=\sqrt{2}\,{\rm sin}(n\pi y)\,$}.
Finally, the $|\Pi|^2$ contribution to 
$\left\langle\delta g\,\delta g\right\rangle_2$ will also be modified
at finite $V/V_c$ as compared to its asymptotics given by
Eq.~(\ref{LK-coefficient}). The corresponding expression\cite{dipl} is
too cumbersome, and we do not reproduce it here. Combining all the
contributions we get the result shown graphically in Fig.~\ref{plot}. 

\section{Subleading terms in the strong-dephasing regime}
\label{app-subleading_terms}

In this Appendix we show that the terms 
$\left\langle\delta g\,\delta g\right\rangle_0$ and
$\left\langle\delta g\,\delta g\right\rangle_1$
remain small compared to the leading term, 
$\left\langle\delta g\,\delta g\right\rangle_2\,$, at high voltages,
\mbox{$V/V_c\gg g\,$}, i.e.~in the regime of strong dephasing.
Let us first analyze the behavior of the contribution (\ref{Xi_1}),
\be
\left\langle\delta g\,\delta g \right\rangle_1 =
32\,V_c\!\!\int\limits_0^{V_1/V_c}\!\!\!dz_1\:\frac{\partial}{\partial
  V_1}\Xi_{z_1-\frac{V_2}{V_c}}\Big|_{V_1=V_2=V}\:,
\ee
which to leading order in $V_c/V$ is
\be
\left\langle\delta g\,\delta g\right\rangle_1 =
32\left[\,\pi V_c\,\frac{\partial}{\partial V_1}\tilde{\Xi}(0)
\Big|_{V_1=V_2}+\int dt\:\tilde{\Xi}(t)\right]\:.
\ee
The last term in the brackets is just twice
$\left\langle\delta g\,\delta g\right\rangle_0\,$,
which we will discuss separately below.
The first term in the brackets gives up to a numerical prefactor
\be
\left\langle\delta g\,\delta g \right\rangle_1
\:\sim\:\left(\frac{V_cg}{V}\right)^{5/4}\:.
\label{intermediate_term}
\ee
Clearly, Eq.~(\ref{intermediate_term}) is not only smaller by a factor
$1/g$ compared to the leading term, Eq.~(\ref{first_term}), at the
lower bound of the strong dephasing regime \mbox{$V/V_c\sim g\,$}, but
also decays faster with increasing voltage.

It remains to show that the contribution of
$\left\langle\delta g\,\delta g\right\rangle_0$ also is
small compared to Eq.~(\ref{first_term}).
Setting \mbox{$V_1=V_2=V$} in Eq.~(\ref{path_integral}), we get
\begin{widetext}
\bea
\Xi_0&=&2\int\limits_0^1 dy_1\,dy_2\,\int\limits_0^\infty
d\tau_1\,d\tau_2\int\limits_{\xi_1(0)=y_2}^{\xi_1(\tau_1)=y_1}{\cal
  D}\xi_1(t_1)\int\limits_{\xi_2(0)=y_2}^{\xi_2(\tau_2)=y_1}{\cal
  D}\xi_2(t_2)\,\times\nonumber\\ 
& &{\rm exp}\left\{-\int\limits_0^{\tau_1}
  dt_1\,\frac{\dot{\xi_1}^2}{4}\:\:-\int\limits_0^{\tau_2}
  dt_2\,\frac{\dot{\xi_2}^2}{4}\:\:-\frac{2V}{V_cg}\,F(y_2)
\big|\tau_1-\tau_2\big|\:\:-\frac{2V}{V_cg}\!\!\!\!\int\limits_0^{{\rm 
min}(\tau_1,\tau_2)}\!\!\!\!dt\:\big|\xi_1-\xi_2\big|\:y_2(1-y_2)\right\}\:. 
\eea
\end{widetext}
Since the bare field fluctuations in the time interval between
$\tau_1$ and $\tau_2$ strongly suppress interference, relevant
contributions are given by paths with $\tau_1$ and $\tau_2$ close to
each other. This allows us to approximate the path integral by two
trajectories propagating for the same time, keeping only the plain
exponential suppression factor to account for the time difference:
\begin{widetext}
\be
\Xi_0=2\sqrt{2}\int\limits_0^1\!dy_1dy_2
\int\limits_0^\infty\!d\tau_1\!\!\int\limits_0^{\tau_1}\!d\tau_2\:{\rm 
  exp}
\left\{-\frac{2V}{V_cg}\,F(y_2)\left|\tau_1-\tau_2\right|\right\}
\int\limits_{\xi(0)=0}^{\xi(\tau_1)=0}\!\!\!\!{\cal
  D}\xi(t_1)\:
{\rm exp}\left\{-\int\limits_0^{\tau_1}\!
  dt\,\left[\frac{\dot{\xi}^2}{4}+
\frac{2V}{V_cg}\sqrt{2}\:|\xi|\,y_2(1-y_2)\right]\right\}\:,
\ee
\end{widetext}
where again we transformed to the sum and the difference of the
coordinates $\xi_1\,$, $\xi_2$ and carried out the trivial path
integral over the sum.
We can also perform the integral over $\tau_2$,
\begin{widetext}
\bea
\Xi_0&=&\sqrt{2}\:\frac{V_cg}{V}
\int\limits_0^1\frac{dy}{F(y)}\:\int\limits_0^\infty
d\tau\left(1-{\rm e}^{-\frac{2V}{V_cg}F(y)\,\tau}\right)
\int\limits_{\xi(0)=0}^{\xi(\tau)=0}\!\!{\cal
  D}\xi(t)\:{\rm exp}\left\{-\int\limits_0^\tau
  dt\left[\frac{\dot{\xi}^2}{4}+
\frac{2V}{V_cg}\sqrt{2}\:|\xi|\,y(1-y)\right]\right\}\nonumber\\
&=&\sqrt{2}\:\frac{V_cg}{V}
\int\limits_0^1\frac{dy}{F(y)}\:\int\limits_0^\infty\frac{d\tau}{\left[ 
\frac{2V}{V_cg}\sqrt{2}\,y(1-y)\right]^{1/3}}
\left(1-{\rm exp}\left\{{}-\frac{\frac{2V}{V_cg}F(y)\,\tau}{\left[
\frac{2V}{V_cg}\sqrt{2}\,y(1-y)\right]^{2/3}}\right\}\right)\,
{\cal I}(0,\tau)\:\:,
\eea
\end{widetext}
where \mbox{$(\partial_\tau-\partial_\eta^2+|\eta|)\,{\cal
    I}(\eta,\tau)=\delta(\eta)\,\delta(\tau)\,$}. 
This integral is dominated by $y$ close to $0$ or $1$,
and should be cut off at \mbox{$y\sim y_c$} and \mbox{$(1-y)\sim y_c\,$},
where $y_c$ is given by Eq.~(\ref{yc}). This yields
\be
\left\langle\delta g\,\delta g\right\rangle_0 
\:\sim\: g\,\frac{V_c}{V}\:\:,
\label{zero_term}
\ee
so that
$\left\langle\delta g\,\delta g\right\rangle_0$ remains
smaller by a factor $\sim 1/g$ than
$\left\langle\delta g\,\delta g\right\rangle_2$
in the strong dephasing regime.  
Since both Eq.~(\ref{intermediate_term})
and Eq.~(\ref{zero_term}) are small compared to Eq.~(\ref{first_term})
at high voltages, we thus have shown that the asymptotic suppression
of the conductance fluctuations is given by Eq.~(\ref{first_term}).


\begin{thebibliography}{1}
\bibitem{lk-note}
Note that the diagrams c-diff and c-coop, as well as some
possibilities of inserting additional impurity lines in the Hikami
boxes, are missing in Fig.~1 of Ref.~\onlinecite{Larkin_Khmelnitskii}.
However, the expression given in Ref.~\onlinecite{Larkin_Khmelnitskii}
appears to give correctly the sum of all diagrams.
\end{thebibliography}
\end{document}